\documentclass[12pt]{iopart}

\usepackage{setstack}

\usepackage{graphicx}
\DeclareGraphicsExtensions{.pdf,.png,.jpg,.eps}

\begin{document}

\title[Measured redshift invariance of photon velocity]{Measured redshift invariance of photon velocity}
\author[Miller, et al.]{J. B. Miller, $^{1}$ T. E. Miller,$^{2}$ M. J. Hoffert,$^{3}$ L. A. Dingle,$^{3}$ R. Harwell$^{3}$ and E. Hayes,$^{3}$ }

\address{$^{1}$ Chemistry Department - MS5413, Western Michigan University, Kalamazoo, MI 49008, USA\ead{john.b.miller@wmich.edu}}
\address{$^{2}$ 108 Titleist Drive, Bluefield, VA 24605, USA\ead{tommy81@alumni.princeton.edu}}
\address{$^{3}$ Tierra Astronomical Institute, 2364 S. Annadel, Rowland Heights, CA 91748, USA\ead{mjhoffert@earthlink.net}}

\begin{abstract}
We report the first direct photon velocity measurements for extragalactic objects.  A fiber-optic, photon time-of-flight instrument, optimized for relatively dim sources ($m\>12$), is used to measure the velocity of visible photons emanating from galaxies and quasars.  Lightspeed is found to be $3.00\pm0.03\times10^{8} \mathrm{m s}^{-1}$, and is invariant, within experimental error, over the range of redshifts measured ($0\leq z\leq1.33$).  This measurement provides additional validation of EinsteinÕs theory of General Relativity (GR) and is consistent with the  Friedmann-Lema\^{i}tre-Robertson-Walker (FLRW) metricl, as well as several alternative cosmological models, notably the hyperbolic anti-de Sitter metric, though not with the pseudo-Euclidean de Sitter metric.  
\end{abstract}

\submitto{\JPG}
\maketitle

\section{Introduction}
A basic tenet of relativity is the constancy of the measured speed of light, regardless of the observerÕs reference frame.  This is so crucial to modern physics, that the speed of light in vacuum, $c_0=2.99792458\times10^8 \mathrm{m s}^{-1}$, besides being one of the most carefully measured of the fundamental constants, is now defined to be exact \cite{mt} and related constants scale accordingly.  However, the vast majority of the published lightspeed measurements have been made using photons that were produced within our Solar system, and, to our knowledge, all published accounts have been based on photon sources within the Milky Way Galaxy. \cite{dart, gkm,nieu,fein,mm03}. Thus the photon travel distances and look-back times for all previous light speed measurements have been negligible on the scale of the universe. That is to say, the redshift, $z$, of the observed photons was effectively zero. In astronomical observations, redshift, $z = \lambda_{obs}/\lambda_{emit}-1$, is the measured increase in the wavelength of a photon measured by an observer ($\lambda_{obs}$) relative to its wavelength when it was emitted ($\lambda_{emit}$).  Redshift is used as an observational proxy for distance, which is an inconvenient quantity to measure directly for most astronomical objects.  No direct measurements of photon velocity from very distant, high-redshift sources have been previously reported.
We thus undertook to determine the velocity of photons cobs originating from high-redshift, quasar sources. In this work we report measurements of the photon velocity from sources at cosmologically significant distances, with redshifts $0\leq z\leq1.33$.  The intent was to test the invariance of light speed between different relativistic reference frames and between current and earlier epochs of the universe.  One goal was to provide solid observational data intended to eliminate or support some of the various hypotheses regarding variation in light speed as a fundamental constant.  A second goal was to provide data that might distinguish between some cosmological models. 
\subsection{Light speed as a fundamental constant}
Measuring the values of fundamental constants during earlier epochs probes for evolution in physical law.  Recently, there have been claims for \cite{wfc,mwf, wmf} and against \cite{scp,lcm} temporal variation of the fine structure constant $\alpha=e^2/4\pi\epsilon_{0}\hbar c$ , and for a small variation of the proton-electron mass ratio $\mu=m_p/m_e$  \cite{rbh}.  Furthermore, a great deal of interest has also been focused on cosmological models with varying speed of light to address the cosmological problems of flatness, horizon, and the cosmological constant \cite{am} or to avoid invoking inflation \cite{mof}.  It has been suggested that arbitrary light speed variations with time, $\dot{c}\left(t\right)\ne0$ , may require breaking covariance or Lorentz invariance \cite{buc}, or violate charge conservation \cite{lsv}, though such objections need not apply in models that contain particle horizons \cite{cjp}.  Some of the theories of current interest, including quantum gravity \cite{witt} and string theory \cite{nie}, incorporate spacetime geometries with inherent event horizons, such as Schwarzchild \cite{mag}, de Sitter \cite{claymoff} and anti-de Sitter, or Brans-Dicke universes\cite{BarrMag} , or even more complex constructs of those geometries as subspaces within various brane topologies \cite{youm,eno}.  Thus the question of variation of fundamental constants may be inextricably linked with the geometry of the universe.
\subsection{Light speed in cosmological models}
The interpretation of the redshift is also dependent on the cosmological model one adopts. In this context, one may pose the na\"{\i}ve question that, if the wavelength of light from redshifted objects is different from ÒnormalÓ, might not other properties, such as the photonÕs velocity, be ÒabnormalÓ as well? The question of the speed of the light emanating from redshifted objects may thus be either trivial or fundamental, depending on oneÕs point of view and degree of conviction in Einstein's theory of General Relativity (GR).  Some may say that we know from GR that any measured lightspeed will be constant, so measurement is futile and the result will be trivial.  However, no amount of theorizing can substitute for an actual measurement.  The photon velocity question is thus fundamental, given that lightspeed forms a basis for other measurements in GR.  It might also be argued that any measurement of photon velocity by a physical observer will be local, and thereby not cosmologically relevant.  However, such a measurement would be neither more nor less local than the measurements of photon redshift.
\subsubsection{FLRW Cosmology}
Currently, the most widely accepted model is described by the Friedmann-Lema\^{i}tre-Robertson-Walker (FLRW) metric, and is interpreted as an expanding Hubble universe.  In addition to its utility as an indicator of distance to the photon source, the redshift also provides a look-back time, presuming that the relevant cosmological parameters (\emph{e.g.}, $H_{0}$, $\Omega_{M}$, $\Omega_{V}$, \emph{etc}. for the FLRW metric) are known.  The speed of light is independent of redshift in the FLRW Hubble universe.
\subsubsection{(anti-) de Sitter Cosmology}
A fundamental driving motivation for this research stemmed from our observation that, assuming a redshift associated with a de Sitter (or anti-de Sitter) universe, the distribution of objectsÕ intrinsic brightnesses is independent of redshift  \cite{mm95,mm10}.  For those unfamiliar with de Sitter geometry, we briefly outline the basic spacetime relationships and their significance.  The de Sitter metric can be written in a pseudo-Euclidean form \cite{deS}:

\begin{equation}       \label{euclidean}
\mathrm{d}s^2  =  - \left( {1 - \frac{{r^2 }}
{{\mathrm{R}^2 }}} \right)^{ - 1} {\mathrm{d}}r^2  -  r^2 \left[ {{\mathrm{d}}\psi ^2  + \sin ^2 \psi {\mathrm{d}}\theta ^2 } \right] + \left( {1 - \frac{{r^2 }}
{{\mathrm{R}^2 }}} \right)c^2{\mathrm{d}}t^2
\end{equation}

where $s$ is the proper time, $r$ is the pseudo-Euclidean radial coordinate, $\theta$ and $\psi$ are angular coordinates, $t$ is the time coordinate, and R is the radius of curvature.  This implies a slowing of lightspeed with increasing redshift.  However, the de Sitter metric can also be written in a hyperbolic form:

\begin{equation}       \label{hyperbolic}
{\mathrm{d}}s^2  = \frac{{ - {\mathrm{d}}h^2  - \mathrm{R}^2 \sinh ^2 \frac{h}
{\mathrm{R}}\left[ {{\mathrm{d}}\psi ^2  + \sin ^2 \psi {\mathrm{d}}\theta ^2 } \right] + c^2{\mathrm{d}}t^2 }}
{{\cosh ^2 \frac{h}
{\mathrm{R}}}}
\end{equation}

where $h$ is a hyperbolic distance coordinate.  This implies constancy of lightspeed at all distances in all directions \cite{deS}.
\section{Experimental}
\subsection{Instrumental Concept}
In one common scheme for measuring lightspeed, the travel time for photons from a chopped high-flux source is determined by extracting and detecting a fraction of the photon stream when the light enters the transit delay, with the remainder detected at the end of the delay.  Another common method is to measure the transit time for photons from a pulsed or modulated source to a detector.  The former works well when there are very large numbers of photons arriving in an essentially continuous (albeit modulated) stream, while the latter works well when one can control the source output.  Both common strategies are essentially equivalent to measuring the velocity of a classical object;  photon flux changes are determined at two times and positions, and the velocity $\mathrm{d}x/\mathrm{d}t$, can be calculated.

Neither of these schemes will work for timing the transits of a stream of relatively-few  randomly-arriving, single photons.  Unfortunately, detection of a photon inherently destroys the photon, since its energy is converted into electronic, vibrational, or nuclear excitation.  Detection is thus effectively a ÒmeasurementÓ of the photon momentum.  This poses problems because our ability to simultaneously measure the photonÕs position and momentum is limited by the Heisenberg Uncertainty Principle.  In short, photon detection precludes determining any velocity information for that photon.  Moreover, if we detected any given photon to know when it entered the delay loop, it would no longer be available for detection at the end of the loop.  As there is nothing in the literature on speed of light measurement for guidance in measuring the velocities of single photons arriving unpredictably, a new strategy was required.

A schematic diagram of our photon time-of-flight instrument is shown in Figure 1.  Superficially, this instrument resembles common bench top equipment for measuring lightspeed.  The basic measurement consists of timing the travel for photons through an optical fiber delay loop.  However, the timing strategy and data collection algorithm that enabled our successful observations of dim astronomical sources are not part of typical lightspeed measurement.  The distant astronomical sources that are the subject of this study provide only a sparse stream of individual photons with unpredictable arrival times.

\begin{figure}
        \center{\includegraphics [width=\linewidth]
         {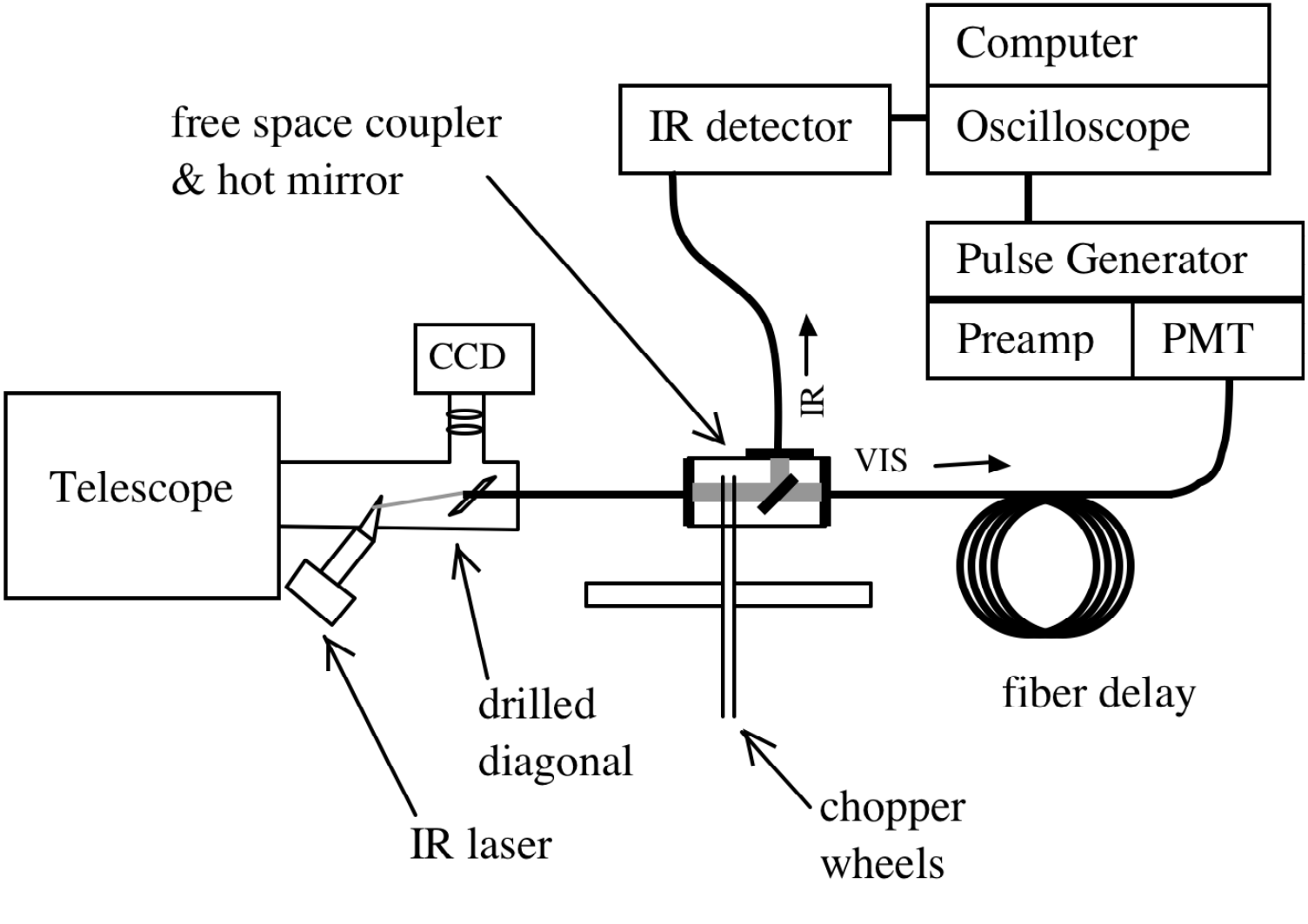}}
	 \caption{ Schematic diagram of the photon time of flight instrument.}
	\label{analysis}
\end{figure}

To achieve the necessary measurements, we constructed a telescopic, dual-band, photon time-of flight instrument based on a very high speed chopper and a fiber-optic delay loop.  One band was a high-flux, near-IR terrestrial source used to determine the chopper transmission function and was used for system timing.  The second band was low-flux visible light from the astronomical sources;  individual photon detection events were co-added after passing through the chopper and the delay loop and used to synthesize a chopper transmission function so that an average velocity of the astronomical photons could be determined. Separation of the bands was achieved by selective transmission and by differences in the IR and visible detector sensitivities.

\subsection{Telescope}
Observations were made under moonless conditions at Manzanita Observatory located at 32$\,^{\circ}$44'5.34"N, 116$\,^{\circ}$20'41.80"W, near Live Oak Springs, California, and operated by the Tierra Astronomical Institute.  The primary telescope was an equatorially mounted 1.22m Ritchey-Chrtien with a 15.24m focal length that had been optically reduced to 5.59m, making the telescope f/4.6.

\begin{table}
	\caption{Astronomical light sources observed}
	\resizebox{\linewidth}{!} {
		\begin{tabular}{l c c c c}
		\hline 
		& \multicolumn{2}{c}{J2000 Coordinates} \\
		Object & RA & Dec & $m$ & $z$ \\
		\hline 
		GSC 282:337 & 12h29m03.2s & +02d03m18s & 13.9 & $\sim$0 \\
		3C273 & 12h29m06.7s & +02d03m09s &12.86 & 0.15834 \\
		HB89 1718+481 & 17h19m38.2s & +48d04m12s & 15.33 & 1.084 \\
		HB89 1634+706 & 16h34m28.9s & +70d31m33s & 14.9 & 1.334 \\
		ÒdarkÓ sky&$\sim$17h19m35s&$\sim$+48d05m00s & $\sim$22-24 & $\sim$0? \\
		\hline
	\end{tabular}
	}
\end{table}

\subsection{Photon Time-of-Flight Instrument}
As shown in Figure 1, the telescope field was focused onto a diagonal mirror that had been diamond micro-drilled to accommodate insertion of an optical fiber.  All fibers were 400 $\mu$m multimode, with 0.22 numerical aperture; this fiber diameter was slightly larger than our estimated seeing diameter of about $350 \mu$m.  Guiding was accomplished using the field of the diagonal, which was projected via a pair of infinity-corrected relay lenses onto a CCD camera (SBIG ST-7).  Since the carefully polished fiber end was in a non-reflective hole and was invisible to the guide camera when illuminated only by light gathered from the telescope, the optimal location for the target objectÕs light was empirically determined by systematic search.  To provide a chopper-timing signal (\emph{vide supra}), the fiber was simultaneously illuminated with a focused infrared laser source (ML725B8F, Mitsubishi, 1310 nm, 10 mW) via an off-axis 30-60-90 prism that was aluminized on the long leg.  The IR laser was found to emit detectable numbers of visible photons, which were eliminated by a long-pass filter (1100 nm cutoff, CVI Laser) before the prism.  Laser light reflected from the diagonal was prevented from reaching the guide camera by a short-pass filter (1050 nm cutoff, Edmund Optics).

The collected target-object photons and the reference IR light were passed via the input fiber to a free-space coupler (FSC), where it was expanded and recombined through a co-linear collimation lens pair (Optics for Research) with a third collimation lens perpendicular to the co-linear pair.  The expanded beam diameter was approximately 4 mm.  A mechanical optical chopper was placed within the FSC.  The chopper consisted of two counter-rotating slotted nickel wheels (1800 slots, 1:1 slot/tab ratio, Scitec, UK) separated by approximately 1.5mm.  The wheels were driven by high-speed, brushless commutator motors (5-75 krpm, Stryker Instruments).  These motors had negligible starting torque and required an external starting mechanism; we used a Dremel$^{\rm TM}$ tool fitted with a rubber wheel.  External cooling was required for extended operation.  The wheels could be stably rotated up to 12 krpm, but were typically operated at 10 krpm, giving a chopping frequency, $\nu$, of \emph{ca.} 600 kHz.  For safe operation, the chopper/FSC assembly was placed in a polycarbonate enclosure, which was evacuated to reduce aerodynamic perturbation of the wheels' rotation.  The chop frequency could be measured independently using the signal from a chopped infrared light source and a universal counter/timer (5334B Hewlett-Packard).  During any given experiment, the chop frequency was constant to within $\pm$3 kHz.  The high chop rate was required to minimize attenuation due to the length of the fiber delay and to maximize the detectability of any photon velocity changes by keeping the sampled time-domain short.
After chopping, the target visible light and the reference infrared light were separated using a 45$\,^{\circ}$ hot-mirror (Edmund Optics).  The chopped IR light beam was (mostly) deflected into the perpendicular collimating lens and refocused onto a 1 m fiber that lead to a fast, amplified GaAs photodiode (TTI-525, Terahertz Technologies).  The visible light was (mostly) transmitted and refocused onto a 159 m fiber-optic delay loop that ultimately illuminated a dry-ice cooled photomultiplier tube (PMT) (9865B, Thorn-EMI) through a final collimating lens.  Careful alignment and iterative optimization of the FSC optics and hot mirror were essential to minimize attenuation of the target light.

The current pulse collected at the anode of the PMT was conditioned with a preamplifier (Nuclear Data) and used to trigger a pulse generator (Stanford Research Systems) acting as a fast discriminator.  To eliminate stray light, the chopper enclosure, the fiber delay housing, and the PMT housing were draped with several layers of rubberized blackout fabric.  The input fiber was surrounded with an opaque, flexible shielding tube.  The telescope attachment was constructed around the very rigid Thorlabs cage system, and any gaps carefully sealed with blackout tape. This portion of the apparatus can be readily mounted on nearly any telescope using common threaded fittings (T-42 or SCT).  The overall increase to the telescope back focus was about 115 mm.
Because the photon arrival rate for the target objects ($<$104 Hz) is much lower than the chop rate ($>$105 Hz), it was useful to have a secondary, high-flux IR timing signal for determining the chopperÕs transmissive state.  Spectrally separating the two signals assures that there is no crosstalk; the photocathode of the PMT is blind to the 1310 nm IR light, while the photodiode is both insensitive and blind to the low-intensity, visible target object light.  The counter-rotation of the wheels results in a precessing Moir\'{e} diffraction pattern (from the overlapping slot arrays) that modulates the beam at the chopper frequency.  The Moir\'{e} pattern is very sensitive to the slightest manufacturing or alignment asymmetry.  We found that the target light and the reference timing light must pass through exactly the same spot on the wheels, at exactly the same angle, to maintain the phase relationship between the target and reference beams.  The target beam and the reference beam must be collinear through the chopper, best achieved in practice by merging the target visible light and the reference IR light at the output of the telescope, where it enters the fiber.
Our method is inherently attenuating; only about 6\% of the photon flux injected into the FSC/chopper assembly survives to pass into the delay loop, which then attenuates the signal by an additional factor of two.  All unnecessary loss points, such as fiber-fiber couplings, have been eliminated.  Other chopping methods were also considered but were discarded as unsuitable from the outset or after initial performance testing.  Commercial microfabricated modulators are generally too low frequency.  Solid state modulators of sufficiently high frequency do not provide adequate beam displacement.  Other rapid modulations methods, such as an electro-optic modulator, or PockelÕs cell, only operate over a narrow bandwidth, and so are effectively even more attenuating that the current system for broad-band emission sources.

\subsection{Detection algorithm and data analysis}
The detection algorithm also merits some attention.  Since timing signal cycles coincident with a photon detection event occur only rarely, and the timing signal period (1.67 $\mu$s) is much shorter than our oscilloscope trigger reset time (\emph{ca}. 150 $\mu$s), it is most efficient to have the oscilloscope trigger on a photon-arrival event, recorded as a (nearly) square pulse with unit height and width $t_{width}$, and then determine its arrival time $\Delta$$t_{trig}$ relative to a trigger point in the reference chop cycle.  Co-adding a large number of n photon events, including nobject events from the target object and $n_{dark}$ dark counts (originating within the PMT) yields a synthetic waveform for the time-integrated PMT signal that represents the probability function

\begin{equation}
\label{prob}
P_{n}\left(t\right)=\frac{1}{n_{object}+n_{dark}}\left(\sum_{i=1}^{n_{object}}f_{i}\left(t\right)+\sum_{i=1}^{n_{dark}}f_{i}\left(t\right)\right)
\end{equation}

comprising short ($t_{width}\leq100$ ns) pulses

\begin{equation}
\label{pulse}
f_{i}\left(t\right)\approx\{
1 \mathrm{\:for\:}\left(t-\Delta t_{trig}\right)\leq t\leq \left(t-\Delta t_{trig}+t_{width}\right); 0 \mathrm{\:for\:all\:other\:}t\}
\end{equation}

The co-added $n_{object}$ events are \emph{de facto} weighted by the chopper transmission function, and produce a modulated signal

\begin{equation}       
\label{chop}
f_{chop}\left(t\right)\approx\cos^{2}\left(\nu\times\left(t+\Delta t_{signal}\right)\right)
\end{equation}

which is offset from the timing signal by the constant difference in signal propagation times of the photodiode and photomultiplier detectors, $\Delta$$t_{signal}$.  In practice, we measure differences in photon travel time using the synthetic PMT waveforms only, and thereby effectively ignore the time offset in $f_{chop}(t)$.  The $n_{dark}$ events occur randomly and show no systematic time dependence within a single chop period, thus they co-add to produce a broad square wave having a period of $1/\nu$.  The signal-to-noise ratio in this experiment is primarily governed by two factors: 1) the total number of detection events, and 2) the relative rates of object and PMT-dark detection events ($\mathrm{d}n_{object}/\mathrm{d}t)/(\mathrm{d}n_{dark}/\mathrm{d}t$).  The precision of the measurements could be improved by increasing the wheel rotation rate, increasing the density of slots (while proportionately decreasing the separation between the wheels), or decreasing the attenuation.  More effective methods would be to increase the number of photon events by using a telescope with larger aperture or integrating for longer times.  Each approach will have its own challenges.
\section{Results and Discussion}
The data in Fig. 2 illustrate the synthetic chopped waveforms used to calibrate the instrument, each created from more than $10^{6}$ individual 100 ns pulses produced by the PMT detection train.  The control light source is an attenuated He-Ne laser emitting 632.8nm photons.  The synthetic chopped PMT waveforms were generated independently using a 1 m and a 159 m delay loop with $dn_{object}/dt$ about $1.4\times10^{4} s^{-1}$ and $6\times10^{3} \mathrm{s}^{-1}$ respectively.  The differing count rates reflect the approximately 3dB attenuation of the longer delay fiber.  The unchopped synthetic PMT dark-count waveform has $dn_{dark}/dt$ of about $20 \mathrm{s}^{-1}$, and was created from only about $2\times10^{5}$ pulses.

\begin{figure}
        \center{\includegraphics [width=\linewidth]
         {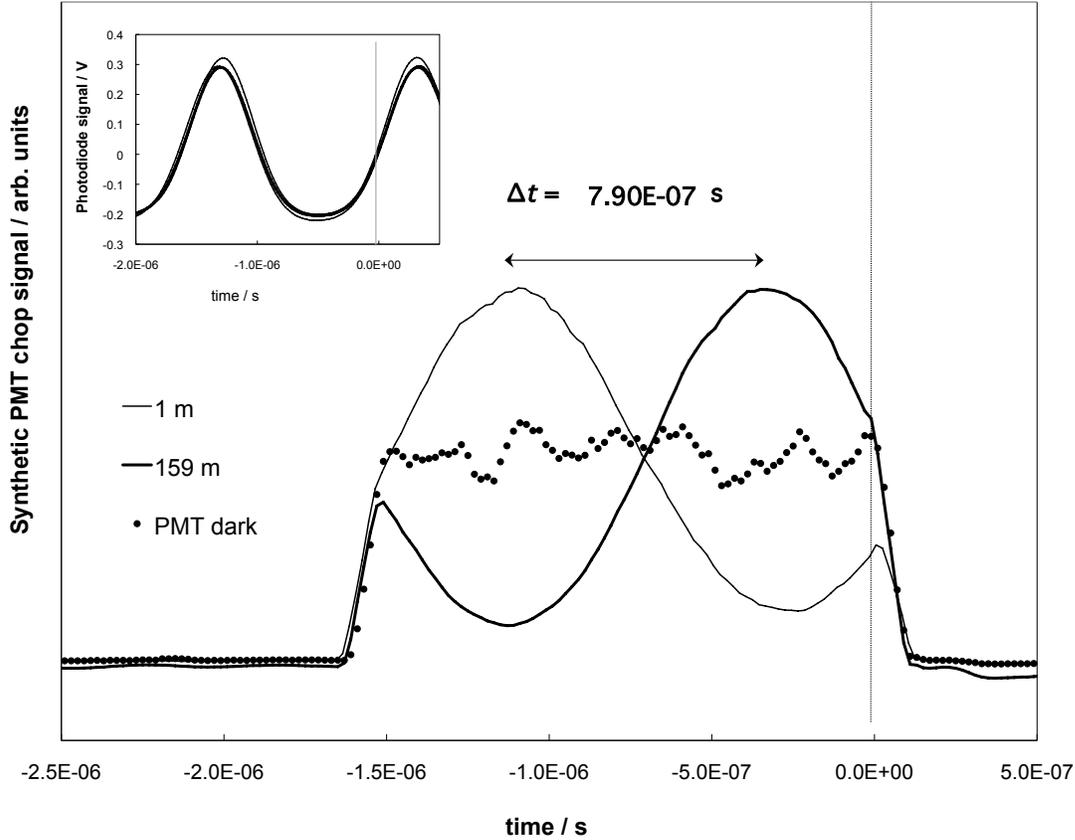}}
	 \caption{ Photon velocity analysis for terrestrial light.  Narrow line Ñ 1 m delay loop; thick line Ñ 159m delay loop; points Ñ PMT dark signal.  The inset graph shows the chopped IR signal.  The gray vertical line in both plots indicates the trigger position for the reference timing signal.}
	\label{analysis}
\end{figure}

The time difference between the two chopped PMT waveforms represents the photon travel time, and is determined by fitting one waveform to the other with a non-linear least-squares algorithm including only data within the full-width at half maximum around the peak of each waveform.  This strategy reduced the need for concern about periodic boundary conditions that are observable in the data, and minimized the impact of differences in the ($\mathrm{d}n_{object}/\mathrm{d}t)/(\mathrm{d}n_{dark}/\mathrm{d}t)$ ratios.  Since terrestrially produced photons are a primary standard for light speed, their travel time was used to calibrate the instrument, obviating the need to physically measure either the length of a long, fragile fiber or its refractive index, which is wavelength dependent.  Instead, from the measured photon transit time, we empirically obtained the product of the fiber length and refractive index (which independently agree with the nominal values for our fiber) to an accuracy of slightly better than 0.5\%; this is the limiting precision for our instrument in its current embodiment.
The data in Figure 3 are the synthetic chopped PMT waveforms for the astronomical light sources listed in Table 1 using the long (159 m) delay.  Each dataset was created from at least $5\times10^{5}$ individual pulses and was normalized using the minimum and maximum values for each waveform, ignoring data that were part of, or within several PMT pulse widths of, the flat minima on the extremes.  In the plot, the curves have been vertically offset for clarity.  The large differences observable in the waveform noise for the various targets reflects the reduction in $(\mathrm{d}n_{object}/\mathrm{d}t)/(\mathrm{d}n_{dark}/\mathrm{d}t)$ as the sources get dimmer.  It was noteworthy that we could observe a slightly modulated PMT signal even from the ÒdarkÓ sky (estimated magnitude 22-24) despite the fact that the count rate was scarcely above that of the PMT dark.  Of course, the quality of that dataset was inadequate to perform a meaningful fit, and the relatively few photon events were most likely due to photons of terrestrial or solar origin in any case.
\begin{figure}
        \center{\includegraphics [width=\linewidth]
         {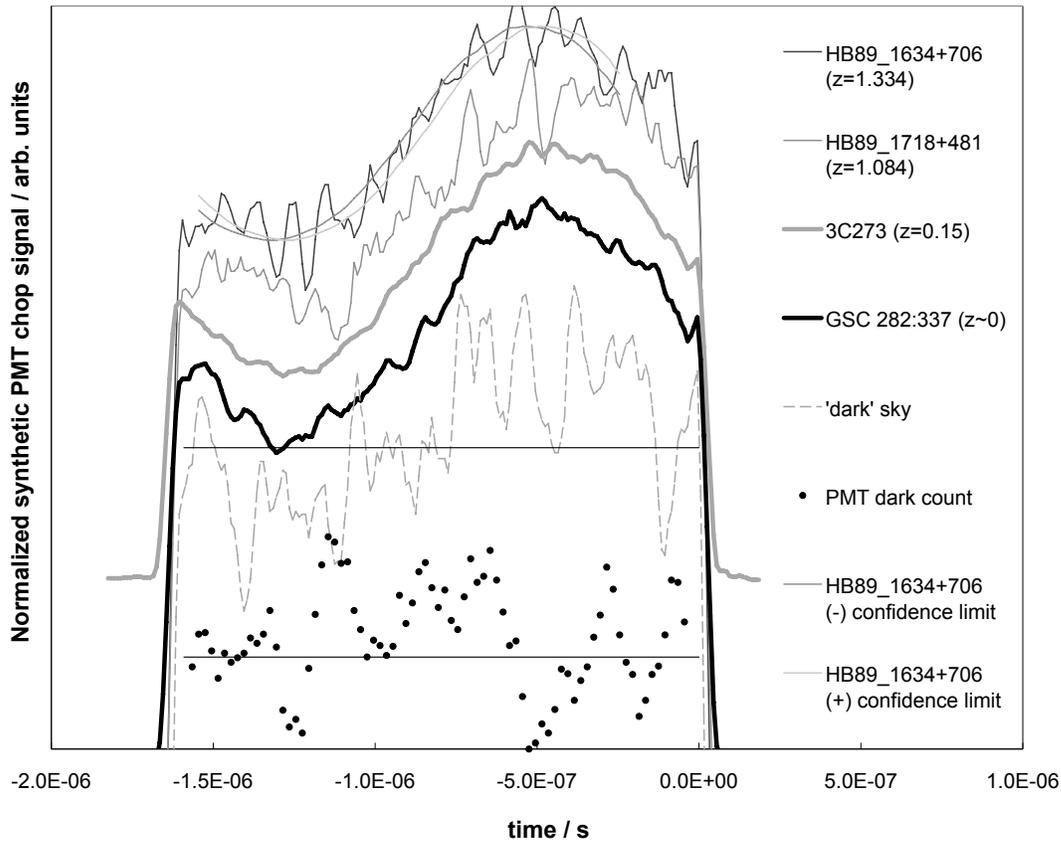}}
	 \caption{ Photon velocity analysis for astronomical light sources using a long (159 m) delay.  The synthetic chopped PMT waveforms were normalized using averaged minimum and maximum values within the period and have been vertically offset for clarity.  The smooth curves superimposed on the HB89 1634+706 data are fits at the 95\% confidence limits.  The horizontal lines were added to guide the eye and highlight the difference between a ÒPMT darkÓ signal and Òdark skyÓ light.}
	\label{analysis}
\end{figure}

There was little qualitative difference between the observed photon velocities of any of the astronomical sources.  Quantitative velocities, shown in Figure 4, were obtained by fitting to the waveform from terrestrial light, analogous to the analysis carried out for calibration. Photon velocities for all objects fall within the range of $3.00\pm0.04\times10^{8} \mathrm{m s}^{-1}$, which is similar to the standard regression error for this data set of $\pm0.03$, though about twice the limiting experimental precision.  The displayed 95\% confidence limits were empirically determined from the measured data sets using the $\chi^{2}$ statistic with one degree of freedom. \emph{Within our experimental uncertainty, we observed no redshift dependence for the measured photon velocities up to $z\leq1.33$.}
\begin{figure}
        \center{\includegraphics [width=\linewidth]
         {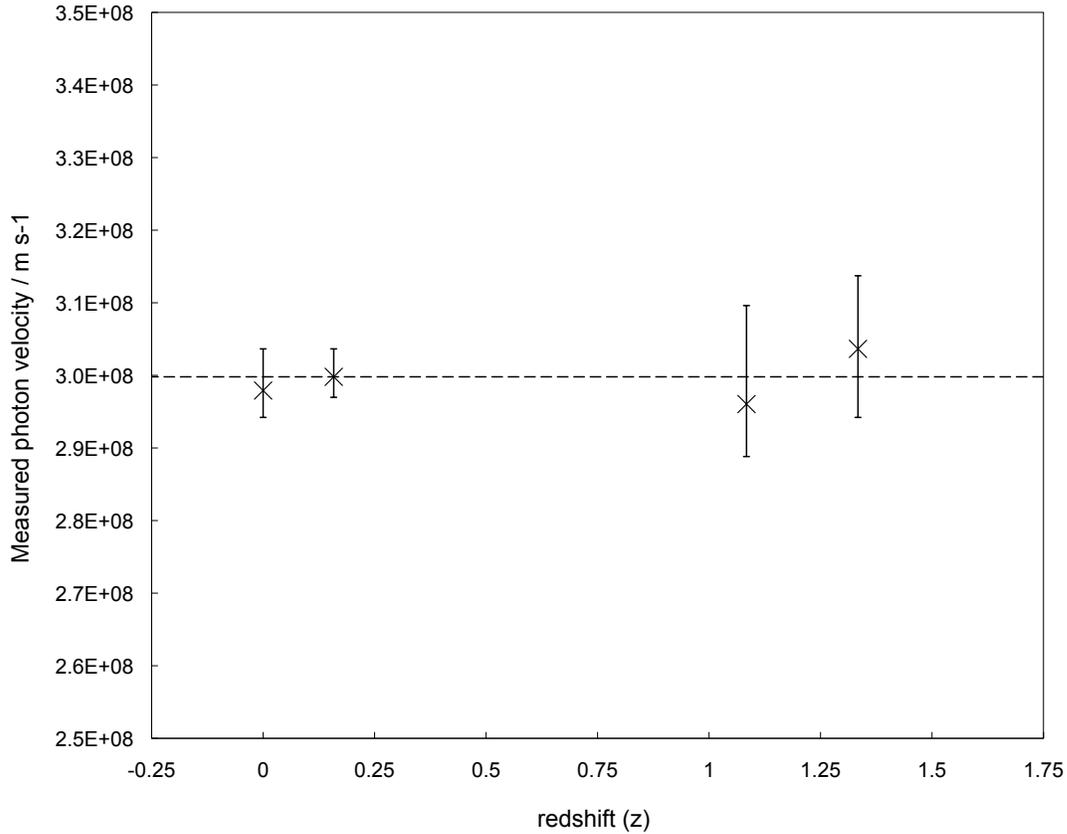}}
	 \caption{ The redshift dependence of the measured photon velocity.  The vertical error bars represent the 95\% confidence limits for each value.  The dashed line is the accepted value for $c$ in vacuum.}
	\label{analysis}
\end{figure}
\section{Conclusion}
The observed invariance of measured lightspeed from extragalactic sources is exactly what one would expect in a universe dominated by General Relativity, thus providing an unequivocal validation of EinsteinÕs theory of General Relativity.  It also suggests that the velocity of photons has been constant, thus constraining the evolution of at least one fundamental constant (c) over a great portion of cosmological time.

The measurement also places some constraints on the cosmologies that could apply to our universe. In particular, if the universe were to have a de Sitter geometry, the redshift invariance of measured lightspeed reported herein is only consistent with the hyperbolic anti-de Sitter metric equation~\Eref{hyperbolic}, not the pseudo-Euclidean form considered elsewhere \cite{mm95}.  The negative curvature of anti-de Sitter space may be consistent with observed phenomena associated with dark matter, such as galactic rotation curves.  Unfortunately, these photon velocity measurements will be inadequate to further distinguish between hyperbolic de Sitter or other various solutions of GR since one can always obtain a coordinate transformation in any given solution for which the speed of light is constant.

While our current measurements lack sufficient precision and look back time to confirm or refute the evolution in physical law suggested by recent observations of proton/electron mass ratio \cite{rbh} they do place significant constraints on many of the variable speed of light theories that have been proposed. The observed redshift-lightspeed invariance virtually eliminates models wherein any lightspeed variation is large, including the power-law light speed variation associated with certain brane world model parameters \cite{Bar}.  Models where there are drastic changes in lightspeed associated with matter-radiation decoupling or sudden changes associated with other brane world assumptions are not addressed by these observations.  Testing these models would require extending these measurements to still-higher redshift, generally dimmer sources, as well as increasing the precision of the current method. 

\ack

This research has made use of the NASA/IPAC Extragalactic Database (NED), which is operated by the Jet Propulsion Laboratory, California Institute of Technology, under contract with the National Aeronautics and Space Administration.  The authors thank Ryan O'Donnell (Stryker Instruments) and Richard Welch (WMU) for significant technical assistance.  The late Dr. Marc Perkovic provided essential equipment loans.  This work was made possible through partial support from Western Michigan University.

\end{document}